\documentclass[11pt,onecolumn,aps,superscriptaddress,
               amsmath,amssymb,nofootinbib]{revtex4}      
\usepackage{graphicx}

\usepackage{graphicx,epsfig}

\usepackage{dcolumn}
\usepackage{bm}

\begin{document}

\begin{flushright}
\baselineskip=12pt \normalsize
{ACT-14-09},
{MIFP-09-54}\\
\smallskip
\end{flushright}

\title{Stringy Neutralino Dark Matter in Light of CDMSII}

\author{James A. Maxin}
\affiliation{George P. and Cynthia W. Mitchell Institute for
Fundamental Physics, Texas A\&M
University,\\ College Station, TX 77843, USA}
\author{Van E. Mayes}
\affiliation{Physics Department, Arizona State University,\\ Tempe, AZ 85287-4111, USA}
\author{D.V. Nanopoulos}
\affiliation{George P. and Cynthia W. Mitchell Institute for
Fundamental Physics, Texas A\&M
University,\\ College Station, TX 77843, USA}
\affiliation{Astroparticle Physics Group, Houston
Advanced Research Center (HARC),
Mitchell Campus,
Woodlands, TX~77381, USA; \\
Academy of Athens,
Division of Natural Sciences, 28~Panepistimiou Avenue, Athens 10679,
Greece}

\begin{abstract}
\begin{center}
{\bf ABSTRACT}
\end{center}
Recently, the CDMS experiment has reported the possible first direct-detection of dark matter.  
We update the direct-detection constraints for neutralino dark matter arising in a realistic string model
constructed from intersecting D6-branes taking into account the recent data from the CDMS collaboration.  We find
that there are well-defined regions of the supersymmetry parameter space where neutralino dark matter satisfying  
the CDMS and all other experimental limits may be obtained. This results in a set of distinct superpartner and Higgs spectra
which may be tested at LHC.    
\end{abstract}

\maketitle

\newpage

\section{Introduction}

Observations in cosmology and astrophysics suggest the presence of a stable dark matter particle.  Recently the Cryogenic Dark Matter Search (CDMS) experiment has announced what may be the first positive 
evidence for the direct detection of WIMP-like dark matter~\cite{Cooley:2009bn}, including the detection of two candidate events, although there is a 23$\%$ chance that these events could be due to background. A natural candidate for WIMP-like dark matter is the lightest supersymmetric partner (LSP)~\cite{Ellis:1983ew} in supersymmetric models which include R-parity conservation, which is usually the lightest neutralino $\widetilde{\chi}_{1}^{0}$~\cite{Ellis:1983ew, Goldberg:1983nd}.  
Limits on the dark matter detection cross-section can be used to constrain the possible superpartner and Higgs spectra which may be observed at the Large Hadron Collider (LHC).  Namely, only those spectra which possess an LSP consistent with the CDMS constraints as well as all other constraints are viable.  The discovery of a neutralino LSP at LHC combined with data from direct and indirect detection experiments may shed 
direct light on the identity of the dark matter.  

Ultimately, supersymmetry breaking and the resulting superpartner spectra should be understood in
terms of a fundamental theory.  At present, the best candidate for such a theory is string theory.  In particular, Type IIA string compactifications involving $D6$-branes intersecting
at angles (and their Type IIB duals including F-theory extensions) have been an exciting direction in which to explore model-building and phenomenology. Such models have been the subject of much study in recent years, and we refer the reader to
\cite{Blumenhagen:2005mu, Blumenhagen:2006ci} for recent reviews.   In contrast to phenomenological frameworks such as mSUGRA, the supersymmetry breaking soft terms
in intersecting $D6$-branes are in general non-universal~\cite{Kors:2003wf}. Thus, it is possible to obtain a different parameter space than what is
typically considered.     

An intersecting D-brane model of this type with realistic features was first constructed in~\cite{Cvetic:2004ui,Chen:2006gd} and studied in~\cite{Chen:2006gd, Chen:2007px, Chen:2007zu}.  In this three-generation Pati-Salam model, it is possible to obtain realistic Yukawa matrices for quarks and leptons, obtain tree-level gauge unification at the string scale, and obtain realistic supersymmetry spectra satisfying all experimental constraints.  The phenomenological consequences of this model at LHC were considered in~\cite{Chen:2007zu} and~\cite{Maxin:2009ez}, and the implications for direct and indirect dark matter detection were studied in~\cite{Maxin:2009qq}.  In the present work, we update these results taking into account the new limit on the dark matter direct detection cross-section from the CDMS collaboration.  Although there is a 23$\%$ chance that the two events detected by CDMS are due to background, it is of interest to entertain the possibility that these events
are real events due to dark matter.  Thus, we find that there are points within the parameter space of this model which are consistent with the CDMS results and which satisfy all experimental constraints.  

\section{WIMP-Nucleon Direct Detection Cross-sections}

We investigate regions of the intersecting $D$6-brane model parameter space that satisfy all the experimental constraints and possess WIMP-nucleon spin-independent cross-sections $\sigma_{SI}$ near the latest CDMS II upper limit of $\sigma_{SI}$ = 3.8 $\times~10^{-8}$ pb at a mass of 70 $GeV/c^{2}$. The results of~\cite{Maxin:2009qq} indicate the regions of the parameter space we wish to target for WIMP-nucleon spin-independent cross-sections near the CDMS II upper limit are regions with low $m_{3/2}$ and high tan$\beta$. Soft-supersymmetry breaking terms are generated using the equations given in~\cite{Chen:2007zu,Maxin:2009ez}. The soft terms are input into {\tt MicrOMEGAs 2.0.7}~\cite{Belanger:2006is} using {\tt SuSpect 2.34}~\cite{Djouadi:2002ze} as a front end to run the soft terms down to the electroweak scale via the Renormalization Group Equations (RGEs) and then to calculate the corresponding relic neutralino density, while $\mu$ is determined by the requirement of radiative electroweak symmetry breaking (REWSB). However, we do take $\mu > 0$ as suggested by the results of $g_{\mu}-2$ for the muon. We use a top quark mass of $m_{t}$ = 173.1 GeV~\cite{:2009ec}. The direct detection cross-sections are calculated using {\tt MicrOMEGAs 2.1}~\cite{Belanger:2008sj}. We employ the following experimental constraints:

\begin{enumerate}

\item The WMAP 2$\sigma$ measurements of the cold dark matter density~\cite{Spergel:2006hy}, 0.095 $\leq \Omega_{\chi} \leq$ 0.129. In addition, we look at the SSC model~\cite{Antoniadis:1988aa}, in which a dilution factor of $\cal{O}$(10) is allowed~\cite{Lahanas:2006hf}, where $\Omega_{\chi} \lesssim$ 1.1. For a discussion of the SSC model within the context of mSUGRA, see~\cite{Dutta:2008ge}. We also investigate another case where a neutralino LSP makes up a subdominant component and employ this possibility by removing the lower bound.

\item The experimental limits on the Flavor Changing Neutral Current (FCNC) process, $b \rightarrow s\gamma$. The results from the Heavy Flavor Averaging Group (HFAG)~\cite{Barberio:2007cr}, in addition to the BABAR, Belle, and CLEO results, are: $Br(b \rightarrow s\gamma) = (355 \pm 24^{+9}_{-10} \pm 3) \times 10^{-6}$. There is also a more recent estimate~\cite{Misiak:2006zs} of $Br(b \rightarrow s\gamma) = (3.15 \pm 0.23) \times 10^{-4}$. For our analysis, we use the limits $2.86 \times 10^{-4} \leq Br(b \rightarrow s\gamma) \leq 4.18 \times 10^{-4}$, where experimental and
theoretical errors are added in quadrature.

\item The anomalous magnetic moment of the muon, $g_{\mu} - 2$. For this analysis we use the 2$\sigma$ level boundaries, $11 \times 10^{-10} < a_{\mu} < 44 \times 10^{-10}$~\cite{Bennett:2004pv}.

\item The process $B_{s}^{0} \rightarrow \mu^+ \mu^-$ where the decay has a $\mbox{tan}^6\beta$ dependence. We take the upper bound to be $Br(B_{s}^{0} \rightarrow \mu^{+}\mu^{-}) < 5.8 \times 10^{-8}$~\cite{:2007kv}.

\item The LEP limit on the lightest CP-even Higgs boson mass, $m_{h} \geq 114$ GeV~\cite{Barate:2003sz}.

\end{enumerate}

\begin{figure}[ht]
	\centering
		\includegraphics[width=0.75\textwidth]{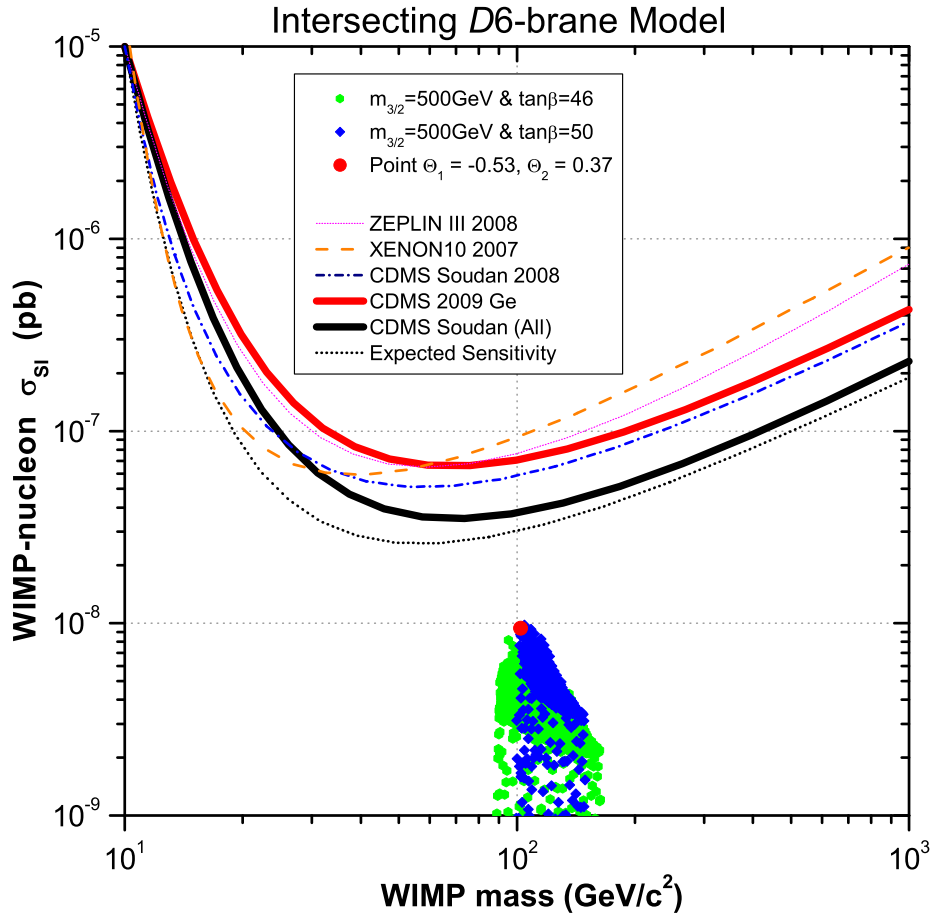}
		\caption{Spin-independent WIMP-nucleon cross-sections of an intersecting $D$6-brane model.  The two cases shown here are $m_{3/2}$ = 500 GeV, tan$\beta$ = 46, and $m_{3/2}$ = 500 GeV, tan$\beta$ = 50. Each marker satisfies all experimental constraints. The point detailed in Table~\ref{tab:masses} and Table~\ref{tab:observables} is annotated on the plot by the red point.}
	\label{fig:sigma_plot}
\end{figure}

We find the optimal cases that are consistent with the latest CDMSII data occur for $m_{3/2}$ = 500 GeV and tan$\beta$ = 46 - 50. The two cases for tan$\beta$ = 46 and tan$\beta$ = 50 are shown in Fig.~\ref{fig:sigma_plot}, where the maximum WIMP-nucleon cross section is $\sigma_{SI} \approx 1 \times 10^{-8}$ pb for these two tan$\beta$. The LSP for these cases is 99.5$\%$ bino.  In addition to the latest data from the CDMSII experiment, the limits resulting from the previous ZEPLIN~\cite{Lebedenko:2008gb}, XENON~\cite{Angle:2007uj}, and CDMS~\cite{Ahmed:2008eu} experiments are also delineated on the plot. For a gravitino mass $m_{3/2} <$ 500 GeV, the LEP Higgs mass constraint of 114 GeV becomes too restrictive, hence, $m_{3/2} \approx$ 500 GeV appears to be most favorable. The points near $\sigma_{SI} \approx 1 \times 10^{-8}$ pb reside in the WMAP 2$\sigma$ relic density region, and these points generate the WMAP observed relic density through stau-neutralino coannihilation. The intersecting $D$6-brane model also possesses regions of the parameter space with chargino-neutralino coannihilation, and furthermore, both of these coannihilation regions can generate very small mass differences between the lightest SUSY particle (LSP) neutralino $\widetilde{\chi}_{1}^{0}$ and the next-to-lightest SUSY particle (NLSP), either the lightest chargino $\widetilde{\chi}_{1}^{\pm}$ or stau $\widetilde{\tau}_{1}$, and thus, a very small relic density. In this scenario, the neutralino can account for only a small portion of the overall composition of the total observed dark matter. The remaining fraction of the observed relic density in this situation would be composed of other particles, such as axions or crytpons, or additional astrophysical matter. In order to enable a direct comparison of Fig.~\ref{fig:sigma_plot} to the data from CDMSII and other direct-detection experiments, for those points in Fig.~\ref{fig:sigma_plot} with a relic density less than the observed WMAP 2$\sigma$ data, we plot a modified WIMP-nucleon cross-section $\sigma_{SI} \times \frac{\Omega_{\chi}}{\Omega_{WMAP}}$. 

The points in Fig.~\ref{fig:sigma_plot} with the largest WIMP-nucleon cross-section $\sigma_{SI}$ possess a WMAP 2$\sigma$ relic density. The SUSY and Higgs spectrum for one of these typical points is shown in Table~\ref{tab:masses} and the low energy observables are given in Table~\ref{tab:observables}. This particular point lives in the stau-neutralino coannihilation region, as evidenced by the less than 15 GeV difference in mass between the LSP $\widetilde{\chi}_{1}^{0}$ and the NLSP $\widetilde{\tau}_{1}$. The lightest CP-even Higgs mass is just above the LEP constraint, and well within the discovery potential at LHC. The process with the largest production differential cross-section at LHC for this point is $q+\overline{q}\rightarrow\widetilde{\chi}_{2}^{0}+\widetilde{\chi}_{1}^{\pm}$. The $\widetilde{\chi}_{2}^{0}$ and $\widetilde{\chi}_{1}^{\pm}$ are virtually degenerate, though the $\widetilde{\tau}_{1}^{\pm}$ is lighter than both the $\widetilde{\chi}_{2}^{0}$ and $\widetilde{\chi}_{1}^{\pm}$, so both the $\widetilde{\chi}_{2}^{0}$ and $\widetilde{\chi}_{1}^{\pm}$ will decay to $\widetilde{\tau}_{1}^{\pm}$ nearly 100\% of the time. The second lightest neutralino will decay to tau through the decay chain $\widetilde{\chi}_{2}^{0}\rightarrow \widetilde{\tau}_{1}^{\pm}\tau^{\mp}\rightarrow\tau^{\pm}\tau^{\mp}\widetilde{\chi}_{1}^{0}$, while the chargino will also decay to tau through $\widetilde{\chi}_{1}^{\pm}\rightarrow \widetilde{\tau}_{1}^{\pm}\nu\rightarrow\tau^{\pm}\nu\widetilde{\chi}_{1}^{0}$, so the end product of both decay chains for this point at LHC will be low energy tau. These points in the parameter space also satisfy all limits on other non-LHC observables such as the limit on the gamma-ray flux from neutralino annihilations.  For further details on the LHC phenomenology of this model, see~\cite{Maxin:2009ez}.

\begin{table}[ht]
	\centering
	\caption{SUSY and Higgs spectrum for a typical point with $\sigma_{SI} \approx 1 \times 10^{-8}$ pb. Here, $m_{3/2}$ = 500 GeV, tan$\beta$ = 50, $\Omega_{\chi}$ = 0.1113, $\Theta_{1}$ = -0.53, and $\Theta_{2}$ = 0.37  (see~\cite{Chen:2007zu} for a complete definition of these parameters). The GUT scale mass parameters for this point are $M3$ = 490 GeV, $M2$ = 160 GeV, $M1$ = 256 GeV, $m_{H}$ = 177 GeV, $m_{L}$ = 446 GeV, $m_{R}$ = 270 GeV, $A_{0}$ = 272 GeV.}
		\begin{tabular}{|c|c||c|c|} \hline
    $Sparticle$&$Mass~(GeV)$&$(S)particle$&$Mass~(GeV)$\\ \hline\hline		
    $\widetilde{\chi}_{1}^{0}$&$102.2$&$\widetilde{t}_{1}$&$853.2$\\ \hline
    $\widetilde{\chi}_{2}^{0}$&$120.1$&$\widetilde{t}_{2}$&$1010.1$\\ \hline
    $\widetilde{\chi}_{3}^{0}$&$661.4$&$\widetilde{u}_{R}$&$1025.7$\\ \hline
    $\widetilde{\chi}_{4}^{0}$&$664.1$&$\widetilde{u}_{L}$&$1083.5$\\ \hline
    $\widetilde{\chi}_{1}^{\pm}$&$120.1$&$\widetilde{b}_{1}$&$928.1$\\ \hline
    $\widetilde{\chi}_{2}^{\pm}$&$666.1$&$\widetilde{b}_{2}$&$1006.8$\\ \hline
    $\widetilde{\tau}_{1}$&$116.5$&$\widetilde{d}_{R}$&$1025.9$\\ \hline
    $\widetilde{\tau}_{2}$&$453.9$&$\widetilde{d}_{L}$&$1086.4$\\ \hline
    $\widetilde{e}_{R}$&$287.3$&$\widetilde{g}$&$1140.8$\\ \hline
    $\widetilde{e}_{L}$&$458.3$&$m_{h}$&$114.1$\\ \hline
    $\widetilde{\nu}_{e/ \mu}$&$451.5$&$m_{A}$&$447.9$\\ \hline
    $\widetilde{\nu}_{\tau}$&$423.9$&$m_{H^{\pm}}$&$455.9$\\ \hline
		\end{tabular}
		\label{tab:masses}
\end{table}

\begin{table}[t]
\caption{Low energy observables for the point $m_{3/2}$ = 500 GeV, tan$\beta$ = 50, $\Omega_{\chi}$ = 0.1113, $\Theta_{1}$ = -0.53, and $\Theta_{2}$ = 0.37 (see~\cite{Chen:2007zu} for a complete definition of these parameters). The GUT scale mass parameters are $M3$ = 490 GeV, $M2$ = 160 GeV, $M1$ = 256 GeV, $m_{H}$ = 177 GeV, $m_{L}$ = 446 GeV, $m_{R}$ = 270 GeV, $A_{0}$ = 272 GeV.}
 \label{tab:observables}
\begin{center}
\begin{tabular}{c c c}
\hline \hline
$\begin{array}{c} g_{\mu} - 2 \end{array}$ &
$\begin{array}{c} Br(b \rightarrow s\gamma) \end{array}$ &
$\begin{array}{c} Br(B_{s}^{0} \rightarrow \mu^{+}\mu^{-}) \end{array}$ \\
\hline
$\begin{array}{c} 41 \times 10^{-10} \end{array}$ &
$\begin{array}{c} 2.98 \times 10^{-4} \end{array}$ &
$\begin{array}{c} 5.3 \times 10^{-8} \end{array}$ \\
\hline \hline
\end{tabular}
\end{center}
\end{table}

\section{Indirect-detection Gamma-ray Flux}

\begin{figure}[t]
	\centering
		\includegraphics[width=0.75\textwidth]{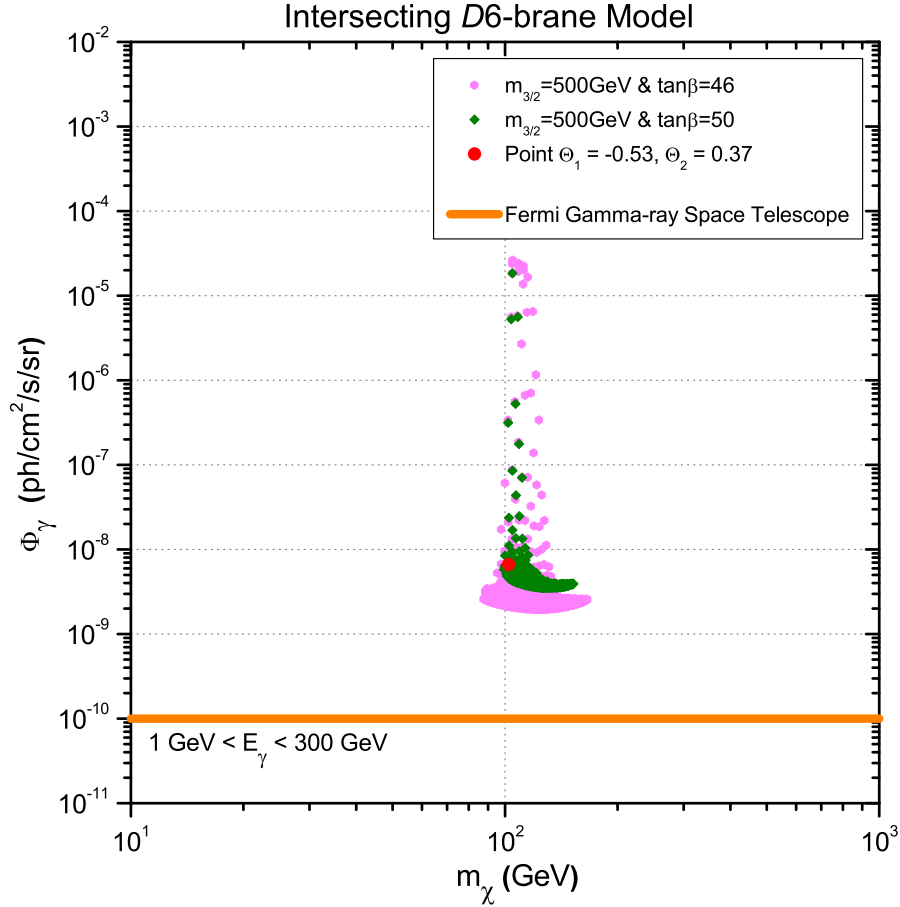}
		\caption{Gamma-ray flux of an intersecting $D$6-brane model with the expected sensitivity of the Fermi Gamma-ray Space Telescope. The two cases shown here are $m_{3/2}$ = 500 GeV, tan$\beta$ = 46, and $m_{3/2}$ = 500 GeV, tan$\beta$ = 50. Each marker satisfies all experimental constraints. The point detailed in Table~\ref{tab:masses} and Table~\ref{tab:observables} is annotated on the plot by the red point.}
	\label{fig:fermi_plot}
\end{figure}

Indirect detection experiments search for high energy neutrinos, gamma-rays, positrons, and anti-protons emanating from neutralino annihilation in the galactic halo and core, or in the case of neutrinos, in the core of the sun or the earth. Here, we focus only on the flux of gamma-rays $\Phi_{\gamma}$ in the galactic core or halo. Two possible decay channels where WIMPs can produce gamma-rays in the galactic core and halo are $\widetilde{\chi}_{1}^{0} \widetilde{\chi}^{0}_{1} \rightarrow \gamma \gamma$ and $\widetilde{\chi}_{1}^{0} \widetilde{\chi}_{1}^{0} \rightarrow q \overline{q} \rightarrow \pi^{0} \rightarrow \gamma\gamma$. The gamma-ray flux $\Phi_{\gamma}$ for an intersecting $D$6-brane model is shown in Fig.~\ref{fig:fermi_plot}, including the expected sensitivity of the Fermi experiment~\cite{Morselli:2002nw}. The gamma-ray flux is calculated using {\tt MicrOMEGAs 2.1}~\cite{Belanger:2008sj}. All regions of the experimentally allowed parameter space for $m_{3/2}$ = 500 GeV, tan$\beta$ = 46 and tan$\beta$ = 50 are within the expected sensitivity of the Fermi telescope.

\section{Conclusion}

The new results from CDMS provide a tantalizing hint that we may be on the verge of the first direct detection of dark matter.  Combined with data
from the now operational LHC as well as indirect detection experiments such as Fermi, it may be possible to finally converge on the fundamental nature of
the dark matter.   In this paper, we studied the implications for dark matter direct and indirect detection experiments for a realistic intersecting D-brane model in light of the new results from CDMS.  We found that there are points in the parameter space which are consistent with the CDMS results, and which satisfy all other experimental constraints.  In addition, we considered the indirect detection gamma-ray flux resulting from neutralino annihilations for these points and find that they are within the sensitivity of the Fermi telescope.   Although the two events detected by CDMS are currently of limited statistical significance,
it is very interesting to entertain the possibility that these events are due to dark matter and see what are the implications of this for different models.  The next few years should be very exciting as more data arrives from CDMS as well as other experiments.  Time will tell whether or not dark matter has finally been directly discovered.  

\section{Acknowledgments}

This research was supported in part by the Mitchell-Heep Chair in High Energy Physics and by DOE grant DE-FG03-95-Er-40917 (DVN and JAM) and by the
U.S. National Science Foundation under grand PHY-0757394.

\end{document}